\documentclass[
  12pt, onecolumn,
]{article}

\title{\bf   Spectral gap  of the totally asymmetric exclusion
       process  at arbitrary filling
}
\author{        O. Golinelli, K. Mallick
\bigskip
\\ \ad          Service de Physique Th\'eorique, 
\\ \ad          Cea Saclay, 91191 Gif-sur-Yvette, France
}
\date{November 19, 2004}

\newcommand  {\ad}{\normalsize\em}      
\usepackage[final]{graphicx}
\newcommand{\figwidth}{\columnwidth}

\newcommand{\Li}{\mathrm{Li}}   

\begin{document}

\maketitle

\begin{abstract}
\normalsize

     We calculate the spectral gap  of the Markov matrix of  the totally
 asymmetric simple  exclusion process (TASEP) on a ring of $L$ sites
 with $N$ particles. Our derivation is simple and  self-contained  
 and extends a previous calculation that
 was valid only for half-filling.  We  use
  a special property of the Bethe equations 
 for TASEP to  reformulate them  as a one-body problem. 
  Our method is closely related to the one used
 to derive  exact large deviation functions of the TASEP.

\medskip \noindent Keywords: ASEP, Bethe Ansatz, Dynamical Exponent,
 Spectral Gap.

\medskip \noindent Pacs numbers: 05.40.-a; 05.60.-k.

\end{abstract}

\section{Introduction}

 The asymmetric simple exclusion process (ASEP) is a  driven
 diffusive system of  particles on a lattice 
 interacting  through  hard-core exclusion, that serves  as a basic  model  
   in various fields  ranging from protein synthesis to traffic
 flow (for a recent review, see  Sch\"utz  2001).  In 
 non-equilibrium statistical physics, the ASEP plays the role of 
   a paradigm thanks to the  variety of phenomenological 
 behavior it displays and to the 
 number  of  exact results  it has led to  in the last decade
 (see,  {\it e.g.},  Derrida 1998). In  particular, 
  ASEP is  an integrable model, {\it i.e.}, the Markov matrix that
 encodes its stochastic dynamics can  be diagonalized by  Bethe Ansatz,
 as  first noticed  by Dhar (1987). Thus, 
  the spectral gap of the Markov matrix, {\it i.e.}, the difference
  between the two  eigenvalues  with largest real parts,  that 
  characterizes the longest relaxation time of the system, 
  can be calculated exactly:   this was first  done 
    for the totally asymmetric simple 
  exclusion process (TASEP)  at half filling
   (Gwa and Spohn 1992) and  later, 
  using  a mapping into the six  vertex model,  Kim (1995)
  treated the case of  the general ASEP at arbitrary filling.  In both works,
 the calculations are  complicated  though the final result for the 
  gap  is fairly   simple. 
  In a recent work (Golinelli and Mallick 2004),   we  presented 
  a concise   derivation of the TASEP gap  at half filling
  that circumvents most of the  technical difficulties  
 thanks to an analytic continuation formula.  However, the 
  half filling condition  seemed  to play a crucial role in our 
 derivation (as well as in the  calculation  of  Gwa and Spohn).
 Our aim in  the present   work is to show that 
 our method can be extended to  the arbitrary filling case.

 We shall study  the TASEP   on a periodic
one-dimensional lattice with $L$ sites (sites $i$ and $L + i$ are
identical). The TASEP is 
 a discrete lattice gas   model on which each lattice site $i$
  ($1 \le i \le L$) is either  empty or occupied by one particle
    ({\em exclusion rule}). Particles evolve according
 to stochastic dynamical rules: a particle on a site $i$ at time $t$ jumps, in
the interval between times $t$ and $t+dt$, with probability $dt$ to the
neighbouring site $i+1$ if this site is empty. 
 As the system is periodic, the  total number 
$N$ of particles is conserved  and  the density 
 (or filling)  is given by  $\rho = {N}/{L}$. 
  A configuration of the TASEP can be
  characterized by the positions of the $N$ particles
on the ring, $(x_1, x_2, \dots, x_N)$ with $1 \le x_1 < x_2 < \dots < x_N
\le L$. If  $\psi_t(x_1,\dots, x_N)$ represents  the probability of this
configuration at time $t$,  the 
evolution of $\psi_t$ is given   by the master equation
 $ {d\psi_t}/ dt = M \psi_t  $,   where  $M$ is the Markov
 matrix.  A right eigenvector $\psi$ is associated with the eigenvalue $E$
 of $M$ if   $ M \psi = E\psi$. 
 Thanks to the Perron-Frobenius theorem, we know that the zero-eigenvalue
 of $M$, which corresponds to the stationary state, is non-degenerate and that
 all the other eigenvalues of $M$ have a strictly  negative real part.
 In the stationary state, all configurations
 have the same probability, given by  $  N! (L-N)! / L!$.

 In the next section, we present  the Bethe  Ansatz  equations
 and restate them as a single self-consistency equation.
 We then calculate the TASEP spectral gap  as a function of the
 density $\rho$ in the limit 
  of  a large system size,  $ L \to \infty$  (Section 3).

\section{The Bethe  Ansatz  Equations}

 The {\em Bethe Ansatz}
assumes that the eigenvectors $\psi$ of $M$ can be written in the form
\begin{equation}
  \psi(x_1,\dots,x_N) = \sum_{\sigma \in \Sigma_N} {\mathcal A}_{\sigma}  \,  
         z_{\sigma(1)}^{x_1} \,  z_{\sigma(2)}^{x_2} \dots z_{\sigma(N)}^{x_N}
  \label{eq:ba} \, , 
\end{equation}
where $\Sigma_N$ is the group of the $N!$ permutations of $N$ indexes. The
coefficients $\{{\mathcal A}_{\sigma}\}$
  and the wave-numbers $\{z_1, \dots, z_N\}$
are complex numbers  determined  by the {\em Bethe equations}. 
In terms of  the fugacity variables $Z_i = 2/z_i -1$, these equations
 become  (Gwa and Spohn 1992)
 \begin{equation}
  (1-Z_i)^N \ (1+Z_i)^{L-N}  
  =  - 2^L \prod_{j=1}^N \frac{Z_j - 1}{Z_j + 1}  \ \ \
  \hbox{with}    \ \ \ i=1,\dots,N     \, .
  \label{eq:bez}   
\end{equation} 
 We note that the right-hand side of these equations is independent
 of the index $i$: this property is true only for the {\it totally}
 asymmetric exclusion process
  and not for the partially  asymmetric exclusion process where  the 
 particles can also jump backwards. 
  Introducing an auxiliary complex variable $Y$,
 the Bethe equations~(\ref{eq:bez})  can be reformulated as explained below 
 (for more details, see Gwa and Spohn
  1992; Golinelli and Mallick 2004).  Consider the one variable 
  polynomial equation  of degree $L$, 
\begin{equation}
   (1-Z)^N \ (1+Z)^{L-N}  = Y \, ,
  \label{eq:Zpoly}   
\end{equation}
and call $(Z_1, Z_2, \dots, Z_L)$ the $L$ roots of this equation.
For a given value of $Y$, the complex numbers  $(Z_1, Z_2, \dots, Z_L)$
belong to a generalized Cassini oval defined by the equation
\begin{equation}
  \left|Z-1 \right|^{\rho}  \,   \left| Z+1\right|^{1 -\rho} = r   \,\,\, 
  \hbox{ with  } \,\,\, r = |Y|^{1/L}   \,  , 
\label{eq:cassini}
\end{equation}
 $\rho$  being the density of the system. The topology of the 
 Cassini oval depends on the value of $r$ (see the figure).  Defining
\begin{equation} 
   r_c =  2 \rho^\rho (1 -\rho)^{ (1 -\rho)}   \, ,
\label{eq:rc}
\end{equation}
we find  that  for $r < r_c$, the locus of the $Z$'s consists
 of two disjoint ovals, with the roots   $(Z_1, Z_2, \dots, Z_N)$
 belonging  to the oval on the right and the 
 roots  $(Z_{N+1}, Z_{N+2}, \dots, Z_L)$
 to the  oval on the left. 
 For $ r = r_c$, the Cassini oval is
 a deformed lemniscate of Bernoulli with a double point at $Z_c = 1 - 2\rho$.
 For  $r > r_c$, the oval is  made of a single loop.  The
 labelling of the    $L$ roots   $(Z_1, \dots, Z_{L})$ 
 is  shown  in  the figure.

\begin{figure}
  \centering
  \includegraphics[width=\figwidth, keepaspectratio]{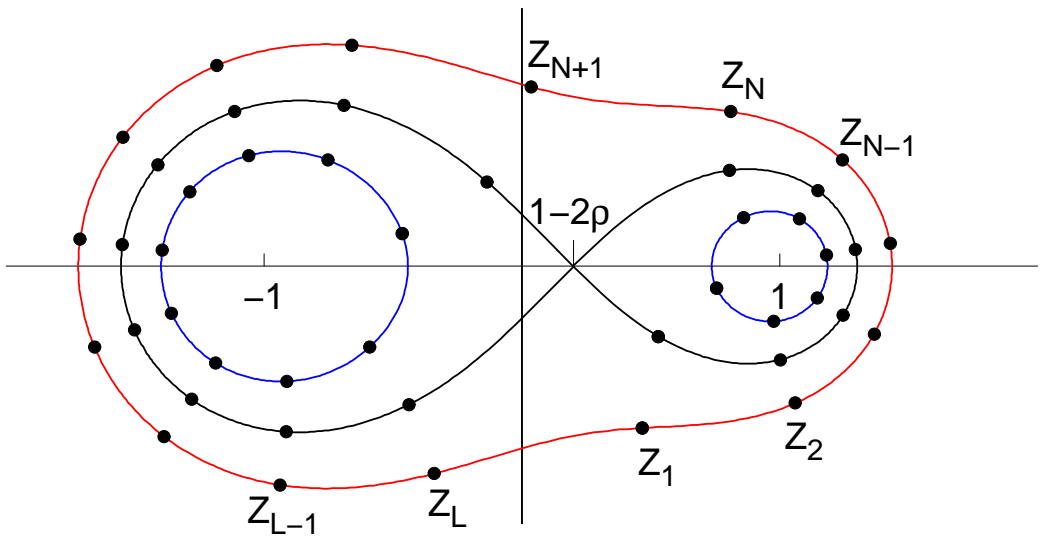}
  \caption{\em 
     Roots  of the equation $(1-Z)^N (1+Z)^{L-N}=Y$.
     Here $N=6$ and $L=15$ (the filling is $\rho = 2/5$) with $Y =
     e^{i\phi} r^L$ for $\phi = \pi/2$; curves are drawn for
       $r/r_c = 0.8$ (blue), $1$ (black)  
   and $1.2$  (red). See text for more explanations. 
   }
  \label{fig}
\end{figure}

 We now introduce a monotonous  function $c: \{1, \dots, N\} \rightarrow
\{1, \dots, L\} $ that 
   selects  $N$ fugacities among
  the  $L$ roots   $(Z_1, \dots, Z_{L})$.
 Defining the function  $A_c(Y)$ as  
\begin{equation}
 A_c(Y) =  -2^L \prod_{j=1}^N \frac{Z_{c(j)} - 1}{Z_{c(j)} + 1} \, ,  
  \label{eq:ac}
\end{equation}
 the Bethe equations become  equivalent to the
 {\it self-consistency} equation 
\begin{equation}
  A_c(Y)  =  Y  \, . \label{eq:ay=y} 
\end{equation} 
Given the choice function $c$ and a solution  $Y$ of this equation, the
$Z_{c(j)}$'s are determined from Eq.~(\ref{eq:Zpoly}) and the corresponding
eigenvalue $E_c$ is given by  
\begin{equation}
 2E_c  =  -N + \sum_{j=1}^N Z_{c(j)} 
  \label{eq:eZ}   \, .  
\end{equation}

 The choice function $c_0(j) = j$ that selects the $N$ fugacities $Z_i$ with
 the largest real parts provides the ground state of the Markov
 matrix. The associated $A$-function and eigenvalue  are given by 
\begin{eqnarray}
  A_0(Y) &=& -2^L \prod_{j=1}^N \frac{Z_j - 1}{Z_j + 1} \, , \label{eq:a0} \\
  2 E_0 &=& -N + \sum_{j=1}^N Z_j  \, .\label{eq:e0}
\end{eqnarray}
  The equation
$A_0(Y) = Y$ has the  unique  solution $Y=0$ that
  yields $Z_j = 1$ for $j \le N $ and
provides the  ground state  with eigenvalue $0$. We shall also need 
the following formula  in the sequel:
\begin{equation}
   \ln  \frac {A_0(Y)}{Y} = -\frac{L}{N} \sum_{j=1}^N 
 \ln \left( \frac{ 1 + Z_j}{2}\right) 
  \label{eq:lna0}   \, .  
\end{equation} 
 To derive this identity, we raise   equation~(\ref{eq:a0})
 to the $N$-th power, use   equation~(\ref{eq:Zpoly}) and take
 the logarithm of the result; however an additional term, which is a discrete
 constant (a multiple of  $2i\pi/N$)  appears in the calculations. 
 This  constant  vanishes 
 identically  because both terms in   equation~(\ref{eq:lna0})
 become real numbers in   the limit $Y \to 0$.

 The spectral gap, given by the 
  first excited eigenvalue,  corresponds to the choice 
  $c_1(j) = j$ for $j = 1, \ldots, N-1$ and $c_1(N) = N+1$
  (Gwa and Spohn 1992). 
 The associated self-consistency  function and eigenvalue 
 are given by (using equations~(\ref{eq:ac}, \ref{eq:eZ}-\ref{eq:e0}))
\begin{eqnarray}
 A_1(Y) &=& A_0(Y) \frac{Z_{N+1}-1}{Z_{N+1}+1} \ \frac{Z_N+1}{Z_N -1} \, ,   
  \label{eq:a1} \\
  2E_1 &=& 2E_0 + (Z_{N+1} -  Z_N)  \, . 
  \label{eq:e1e0}
\end{eqnarray}
 Thus, from  equation~(\ref{eq:a1}), 
 the self-consistency equation that determines the gap reads
\begin{equation} 
   0 =  \ln  \frac {A_1(Y)}{Y} =  \ln  \frac {A_0(Y)}{Y} 
  - \ln \left(
 \frac{ 1 - Z_N}{ 1+ Z_N} \ \frac{ 1 + Z_{N+1}}{ 1 -Z_{N+1}} \right)  \, .   
 \label{eq:sca1}
\end{equation}

  The excitation corresponding to the choice function 
$c(j) = j+1$ for $j = 1, \ldots, N-1$ and $c(N) = L$,  leads 
to the complex-conjugate eigenvalue $\bar{E_1}$.
The eigenvalue  $E_1$ corresponding to 
  the  first excited state is obtained by
 solving the self-consistency equation $A_1(Y) = Y$ where 
 $A_1$ is defined in equation~(\ref{eq:a1}). We  shall solve this 
 equation in the limit $N, L \to \infty$, keeping $\rho = N/L$ constant.

\section{Calculation of the first excited state}

 As in (Golinelli and Mallick 2004),  we start with 
  the   Taylor expansions,   
  of   $A_0 (Y)$  and $E_0 (Y)$ in the vicinity of $Y=0$
   and valid for arbitrary  values of $N$ and $L$.    
In our previous work, we  considered only the half-filling case, 
$L =2N$, for which  the  polynomial equation~(\ref{eq:Zpoly}) reduces to 
$( 1 - Z^2)^N = Y$ and  can be solved {\it explicitly} in terms of Y 
 yielding  the  Taylor series of   $A_0$ and $E_0$. 
  An explicit solution of equation~(\ref{eq:Zpoly}) can not, however, 
 be obtained for  an arbitrary density $\rho$. 
 This  major technical difficulty can be circumvented 
 thanks  to  a contour integral representation  similar to that 
 used for  calculating  large deviation functions  
(Derrida and Lebowitz 1998; Derrida and Appert 1999; Derrida and Evans 1999).

 When $Y \to  0$, the $N$ roots $(Z_1, \dots, Z_N)$ 
 of  equation~(\ref{eq:Zpoly})
  with the  largest real parts
 converge to +1, whereas the $L -N$ remaining roots converge  to -1. 
  We now consider a  positively oriented  contour $\gamma$ 
  that encircles +1 such that
 for  sufficiently small values of 
  $Y$   the roots  $(Z_1, \dots, Z_N)$ lie  inside  $\gamma$
   and   $(Z_{N+1}, \dots, Z_L)$ are  outside $\gamma$. Let
 $h(Z)$ be a function  that is  analytic in a domain  containing  
  the contour  $\gamma$.  We also  define  
\begin{equation}
  P(Z) = (1-Z)^N \ (1+Z)^{L-N}.
\end{equation}
 Because,  by definition, the $Z_j$ are the zeroes of $P(Z) = Y$,
 we obtain,  from the residues theorem, 
\begin{equation} 
   \sum_{m=1}^N  h(Z_m) = \frac{1}{2i\pi} \oint_{\gamma}
 \frac{ P'(Z)}
  { P(Z)  - Y } h(Z)  {\rm d}Z \, .
\label{eq:residus}
\end{equation}
Expanding  the denominator in the contour integral  for small values of $Y$
 (thanks to the formula $(P - Y)^{-1} =  \sum_{k=0}^\infty Y^k/P^{k+1}$
 valid for $ |Y| <|P|$), 
 we obtain the Taylor expansion  
\begin{eqnarray} 
   \sum_{m=1}^N  h(Z_m) &=& \frac{1}{2i\pi} \sum_{k=0}^\infty Y^k  
  \oint_{\gamma}  
  \frac{ P'(Z) }
  {  P^{k+1}(Z)  } h(Z) \,  {\rm d}Z \nonumber \\
  &=&    \frac{1}{2i\pi} \oint_{\gamma}  
  \frac{ P'(Z) } {  P(Z)  } h(Z) \,  {\rm d}Z  
   +    \frac{1}{2i\pi} \sum_{k=1}^\infty \frac{ Y^k }{k}
   \oint_{\gamma}     \frac{  h'(Z) }
  { P^{k}(Z)    }  \,  {\rm d}Z   \,    \nonumber \\
 &=&  N h(1) +  \frac{1}{2i\pi} \sum_{k=1}^\infty \frac{ Y^k }{k}
   \oint_{\gamma}     \frac{  h'(Z)  \, {\rm d}Z  }
  { (1 -Z)^{kN} (1 + Z)^{k(L-N)}  }  \,    \, . 
\label{eq:Taylorh}
\end{eqnarray}
  The second  equality is derived  by  integrating  by parts 
   the terms with  $k \ge 1$; 
  the term $ N h(1)$ in the   third equality  
   is obtained from  the residues theorem. 
  We shall  need the following identity,    valid for any positive integers
 $P$ and $Q$:
\begin{equation}
\frac{1}{2i\pi}  \oint_{\gamma}    
 \frac{ 1}  { (1 -Z)^{P} (1 +Z)^{Q}} \,  {\rm d}Z  =
- {2^{ 1-P-Q}} {\left(\begin{array}{c} P+Q-2 \\ P-1 \end{array} \right) }   
 \, .  \label{eq:ident} 
 \end{equation}

  Using  equation~(\ref{eq:Taylorh}) with 
 $h(Z) = \ln  \left( \frac{1 + Z}{2} \right)$,    we obtain
 from equations~(\ref{eq:lna0}) and~(\ref{eq:ident})
\begin{equation}
\ln  \frac {A_0(Y)}{Y} =   \sum_{k=1}^\infty 
 \left(\begin{array}{c} kL \\ kN \end{array} \right)
   \frac{Y^k}{ k 2^{kL}}
  \, .  \label{eq:Taylora0} 
 \end{equation}
 Similarly,  using   equation~(\ref{eq:Taylorh})
  with  $h(Z) = Z-1 $, we obtain
  from equations~(\ref{eq:e0}) and~(\ref{eq:ident})
\begin{equation}
 2 E_0   =  -   \sum_{k=1}^\infty
 \left(\begin{array}{c} kL-2\\ kN-1\end{array}\right)
 \frac{Y^k}{ k 2^{kL-1}} 
   \, .\label{eq:Taylore0}
\end{equation}

 In the  limit $ L \to \infty$ and  with $\rho$ fixed,
  we obtain  from   the Stirling formula
\begin{equation} 
 \ln  \frac {A_0(Y)}{Y}  
    \to \frac{1}{\sqrt{2\pi \rho(1 -\rho) L}} 
 \ \Li_{3/2} \left(\frac{Y}{r_c^L} \right)
  \, ,   \label{eq:lhsa1}
\end{equation}
 where $r_c$ was defined in equation~(\ref{eq:rc}) and the  
 {\em polylogarithm} function of index $s$, $\Li_{s}$, is given by
\begin{equation}
  \Li_s(z) = \frac{z}{\Gamma(s)} \ \int_0^{\infty} \frac{t^{s-1} \
  dt}{e^t -z} = \sum_{k=1}^{\infty} \frac{z^k}{k^s}  \, .  
  \label{eq:Li}
\end{equation} 
 The function $\Li_{s}$ is defined by the first
 equality on   the whole
 complex plane with a branch cut along the real semi-axis $[1,+\infty)$;  
 the second  equality is   valid only for $|z| < 1$.

 The limit found
 in equation~(\ref{eq:lhsa1}) suggests that  $Y$   can  be  
 parameterized  as follows:
  \begin{equation}
 Y = - r_c^L  e^{u\pi} \, , 
 \label{eq:u} 
\end{equation}
 $u$ being a complex number with $ -1 \le \mathrm{Im}(u) < 1$ 
 and which  remains 
 finite in  the limit $L \to \infty$. Thus $|Y|^{1/L} \simeq r_c$
 and roots of the type  $Z_k$, $Z_{L-k}$ and $Z_{ N \pm k}$
 where $k$ is a fixed  positive integer  are close to  
 the lemniscate double-point $Z_c = 1 - 2\rho$. Writing 
 $Z = 1 - 2\rho + 2 \xi$, with  $\xi \ll 1$, 
 and taking the logarithm of equation~(\ref{eq:bez})
 we obtain 
\begin{equation}
 ( 1 - \rho) \ln \left( 1 + \frac{\xi}{ 1 -\rho} \right)
 +  \rho  \ln \left( 1 -  \frac{\xi}{\rho} \right) 
 = \pi \frac{u +i (2k -1)}{L} \, .
  \end{equation}
  For  fixed  $k$, the   values   $k \le 0$ lead
  to $Z_{N -k}$ and $Z_{L-k} \, $,  and 
 $k  \ge 1$  lead to $Z_k$ and $Z_{N+k}$. 
  Neglecting terms of order 
 $ {\mathcal O}\left( {L^{-3/2}} \right)$, we have 
 \begin{equation}
  Z_N    =   1 - 2\rho + 2i \sqrt{\frac{2\pi \rho(1 -\rho)}{L}} 
   \ (u- i)^{1/2}  + \frac{4\pi}{3} \frac{ 1 -2 \rho}{L}(u- i)
           +\ldots   
  \label{eq:devZn} 
 \end{equation}
 \begin{equation}
 Z_{N+1} =  1 - 2\rho + 2i \sqrt{\frac{2\pi \rho(1 -\rho)}{L}} 
   \ (u + i)^{1/2}  +  \frac{4\pi}{3} \frac{1- 2 \rho}{L}(u +  i)
             +\ldots    \label{eq:devZnp1}
\end{equation}
 Substituting 
 equations~(\ref{eq:lhsa1}),~(\ref{eq:devZn}) and~(\ref{eq:devZnp1})
 in equation~(\ref{eq:sca1}), we obtain the large $L$ limit of the
 Bethe equations for the gap  which, at  the leading order, reads as  
\begin{equation}
  \Li_{3/2}(-e^{u\pi}) = 2i\pi \left[ (u+i)^{1/2} -  (u-i)^{1/2} \right] .
  \label{eq:li32}
\end{equation}
 This equation is the same as that  obtained in  
  (Golinelli and Mallick 2004) and its solution is given by:
\begin{equation}
  u = 1.119 \, 068 \, 802 \, 804 \, 474 \dots 
\label{eq:solu}
\end{equation}
 We can now calculate the eigenvalue corresponding to the
 first excited state.
>From equations~(\ref{eq:e1e0}),~(\ref{eq:sca1}) and~(\ref{eq:Taylore0}),
 we obtain 
\begin{equation}
 2 E_1 =  (2 E_0  + Z_{N+1} - Z_N)  + 
  2 \rho ( 1 - \rho)\ln  \frac {A_1(Y)}{Y}
 =   \label{eq:Taylore1} 
 \end{equation}
$$
  -    \sum_{k=1}^\infty
\frac{ \rho ( 1 - \rho)\left(\begin{array}{c} kL\\ kN\end{array}\right)Y^k  }
 { k (kL-1)2^{kL-1}} 
 + (Z_{N+1} - Z_N) -  2 \rho ( 1 - \rho)
\ln \frac{ (1 - Z_N)(1 + Z_{N+1})}{ (1+ Z_N)(1 -Z_{N+1})} 
\, .        \nonumber  $$

The large $L$ limit of this expression is found  by using  the Stirling
 formula, the expansions~(\ref{eq:devZn}) and (\ref{eq:devZnp1}),
 and the parameterization of $Y$ given by  equation~(\ref{eq:u}).
  We  thus obtain
\begin{eqnarray}
 2 E_1 =  &&  - \frac{1}{L^{3/2}} 
 \sqrt{\frac{2 \rho(1 -\rho)}{\pi}} \left(  \Li_{5/2}(-e^{u\pi})
  - \frac{ 4 \pi^2}{3}i \left[ (u+i)^{3/2} -  (u-i)^{3/2} \right] \right)
         \nonumber \\  && + \frac{4 i \pi }{L} ( 1 -  2 \rho) \, ,
 \end{eqnarray}
 where $u$ is the solution of the  equation~(\ref{eq:li32}), 
 its  numerical value
  being given in equation~(\ref{eq:solu}). Comparing with the value
         $E_{1, \,  \rho =  {1}/{2}}$ 
  obtained at half-filling 
   we conclude that
\begin{equation}
  E_1   =   2 \sqrt{ {\rho( 1 -\rho)} } \,\, 
  E_{1,\, \rho =  {1}/{2}} 
   + \frac{2i\pi}{L}( 1 - 2\rho) 
   \, . \nonumber
\end{equation}
 This equation agrees with the one derived by  Kim (1995). We notice that
  this eigenvalue
 has a non-vanishing imaginary part when 
  the density is different from one half.

\section{Conclusion}

 The asymmetric exclusion process can be mapped into the six vertex model
 and is  thus an integrable model that  can be solved  by 
 Bethe Ansatz. Kim (1995) has used this technique to calculate 
 gaps and crossover functions for the generic  asymmetric exclusion process
  but the   calculations are very complicated.  However, for the
 totally asymmetric exclusion process  
 the  Bethe equations can be reduced to a single
  polynomial equation and their analysis becomes much simpler 
  as shown in the present work.  We  
 have calculated the gap of the TASEP for  arbitrary filling.
  Our method  is closely related  to that  used to calculate  large deviation
 functions (see Derrida 1998, for a review) and  leads to 
  derivations  that  are far 
 simpler than  the ones  presented in previous works. This technique
  can be generalized to calculate any finite excitation close to the ground
 state of the system.

\section*{References}
\begin{itemize}

\item 
 B. Derrida, 1998, 
{\em An exactly soluble non-equilibrium system: the asymmetric simple
 exclusion process},  
 Phys. Rep.  {\bf 301}, 65.

\item 
  B. Derrida, C. Appert, 1999,
{\em Universal large-deviation function of the Kardar-Parisi-Zhang
 equation in one dimension},  
 J. Stat. Phys. {\bf 94}, 1. 

\item 
 B. Derrida, M. R. Evans, 1999,
 {\em Bethe Ansatz solution for  a defect particle in 
the asymmetric exclusion process}, 
 J. Phys. A: Math. Gen. {\bf 32}, 4833.

\item 
 B. Derrida, J. L. Lebowitz, 1998,
{\em Exact  large deviation function in the asymmetric exclusion process},  
 Phys. Rev. Lett.  {\bf 80}, 209.

\item 
  D. Dhar, 1987,
{\em    An exactly solved model for interfacial growth},
  Phase Transitions {\bf 9}, 51.

\item 
O. Golinelli, K. Mallick, 2004,
 {\em Bethe Ansatz calculation of the spectral gap of the
 asymmetric  exclusion process},  
 J. Phys. A: Math. Gen.  {\bf 37}, 3321.

\item 
 L.-H. Gwa, H. Spohn, 1992,
  {\em Bethe solution for the dynamical-scaling exponent of the noisy
  Burgers equation},
  Phys. Rev. A {\bf 46}, 844.

\item 
 T. Halpin-Healy, Y.-C.~Zhang, 1995, 
{\em Kinetic roughening phenomena, stochastic growth, directed polymers and
 all that},
Phys. Rep.  {\bf 254}, 215.

\item 
  D. Kim, 1995,
 {\em Bethe Ansatz solution  for crossover scaling functions
 of the asymmetric XXZ chain  and the Kardar-Parisi-Zhang-type
 growth model},
 Phys. Rev. E {\bf 52}, 3512.

\item 
 G.~M.~Sch\"utz, 2001,
 {Phase Transitions and Critical Phenomena, Vol. 19},
   (Academic, London).

\end{itemize}

\end{document}